\documentclass[showpacs,preprintnumbers]{revtex4}
\usepackage{amssymb}
\usepackage{amsmath}
\usepackage{graphicx}
\usepackage{dcolumn}
\usepackage{bm}
\usepackage{epsf}

\newcommand{\nuc}[2]{\ensuremath{\rm{^{#1}}#2}}  

\begin{document}

\preprint{APS/123-QED}

\title{Low energy $n-\nuc{3}{H}$ scattering:
a novel testground for nuclear interaction}
\author{R. Lazauskas}
\affiliation{DPTA/Service de Physique Nucl\'eaire, CEA/DAM Ile de France, BP 12,
           F-91680 Bruy\`eres-le-Ch\^atel, France}
\author{J. Carbonell}
\affiliation{L.P.S.C.,  53 Av. des Martyrs, 38026 Grenoble, France}
\author{A. C. Fonseca}
\affiliation{Centro de F\'\i sica Nuclear da
Universidade de Lisboa, Av. Prof. Gama Pinto, 2, 1649-003 Lisboa,
Portugal}
\author{M. Viviani, A. Kievsky and S. Rosati}
\affiliation{INFN, Sezione di Pisa and Physics Department, University of Pisa, Pisa I-56100, Italy}
 \date{\today }

\begin{abstract}
The low energy $n-\nuc{3}{H}$ elastic cross sections near the
resonance peak are calculated by solving the 4-nucleon problem
with realistic NN interactions. Three different methods -- Alt,
Grassberger and Shandas (AGS), Hyperspherical Harmonics and
Faddeev-Yakubovsky -- have been used and their respective results are
compared. We conclude on a failure of the existing  NN forces to
reproduce the $n-\nuc{3}{H}$ total cross section.
\end{abstract}
\pacs{21.45.+v, 11.80.Jy, 25.40.-h, 25.10.+s}
\maketitle

\section{Introduction}

\begin{figure}[h!]
\begin{center}\epsfxsize=12.0cm\mbox{\epsffile{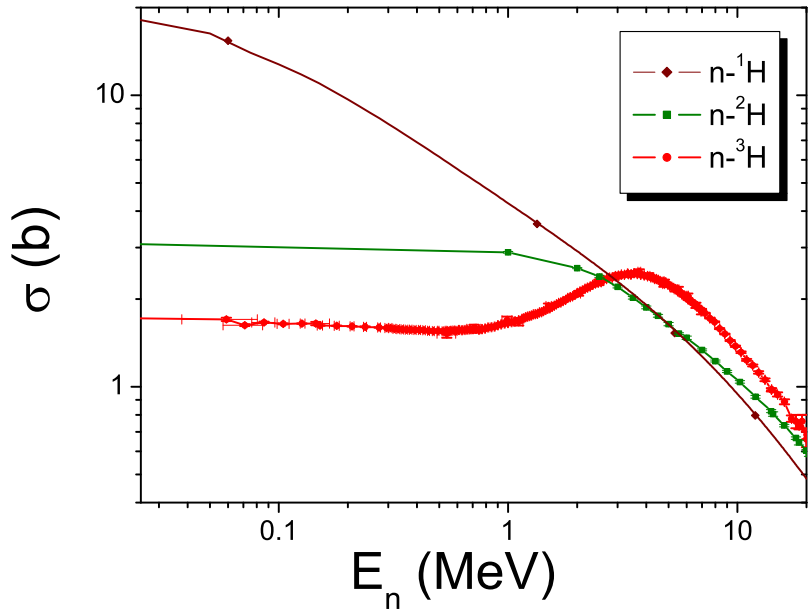}}
\vspace{-1.cm} \caption{Comparison between n-p, n-d and
$n-\nuc{3}{H}$ cross sections for neutron energies in laboratory
frame.} \label{Diff_234N}\end{center}
\end{figure}

The four nucleon (4N) system represents a qualitative jump in
complexity relative to the $A=3$ case, as has been put
forward in several
papers~\cite{CCG_98,CC_98,CCG_99,CCGF_98,Viv98,AF_99,NKG00,Viv01,AF_02}.
It becomes already obvious when comparing the experimental $n-p$,
$n-d$ and
$n-\nuc{3}{H}$  cross sections  displayed in Fig.~(\ref{Diff_234N});
the smooth behaviour of $A=2$ and $A=3$ curves contrasts with
the non trivial structure manifested in $A=4$. This structure, in the energy region
around E$_{cm}\approx$3 MeV, is commonly
associated with negative parity, isospin $T=1$ resonance states \cite{TWH92}. Actually,
$A=4$ is the smallest nuclear system which exhibit characteristic
nuclear properties -- such as saturation in the binding energy --
and in which the simplest nuclear ($n+\nuc{3}{He}\rightarrow
p+\nuc{3}{H}$) and fusion ($d+d\rightarrow n+\nuc{3}{He}$)
reactions take place. Moreover, many reactions involving four
nucleons, like $p + \nuc{3}{He}
\rightarrow \nuc{4}{He} + \nu_{e} + e^{+}$ (the $hep$ process), are
of extreme astrophysical interest, as they play important roles in
solar models and big-bang nucleosynthesis;  the $hep$ process,
for instance, is the source of the highest energy neutrinos from
the Sun.

The study of the 4N  system is particularly interesting as a
``theoretical laboratory" to test new models of the nuclear force.
Unlike the $A=3$ systems, $A=4$ shows a delicate and rich
structure of excited states in the continuum~\cite{TWH92} whose
position and width depends critically on the underlying
nucleon-nucleon (NN) interaction~\cite{AF_02}. In fact, the effect
of (i) the NN P-wave and of (ii) the three-nucleon (3N) force  are
believed to be larger than in the $A=2$ or $3$ systems. Moreover,
it is the simplest system where the 3N interaction in the channels
of total isospin $T=3/2$ can be studied. Therefore it is of the
utmost importance to have reliable few-body techniques powerful
enough to deal with this system. Ever increasing computer power,
development of novel numerical methods, and significant
refinements of well-established techniques have come together to
show that the solution of the four-nucleon bound state problem
with realistic Hamiltonians is reliable.

In Ref.~\cite{Kea01} the binding
energies and other properties of the
$\alpha$-particle were studied using the AV8$'$~\cite{AV18+} NN
interaction; several different techniques
\cite{KG92,KKF89,KK90,VS95,SV98,VKR_04,VKR_95,C88,Wea00,NB99,NVB00,BLO00,Nogga_alpha}
produce results in very close agreement with each other
(at the level of less than 1\%). More recently the Faddeev--Yakubovsky
(FY) equations~\cite{Y67,Nogga_alpha} were solved for different
realistic NN + 3N force models. These calculations are fully
converged in terms of NN partial waves and the results agree with
identical works using the Green's function Monte Carlo
(GFMC)~\cite{Wea00} or Hypherspherical Harmonics (HH)
methods~\cite{VKR_04}.

In this paper, we consider the problem of solving the $4N$
scattering problem with realistic Hamiltonians. As discussed
before, the $A=4$ system is very rich and several elastic and
rearrangement channels are indeed possible already at low energy,
depending on the total charge ($n-\nuc{3}{H}$,
$p-\nuc{3}{H}\leftrightarrow n-\nuc{3}{He}\leftrightarrow d-d$,
$p- \nuc{3}{He}$). Among them, the $n-\nuc{3}{H}$ is the simplest
one: free from Coulomb interaction and, to a very good
approximation, a pure isospin $T=1$ state. However, it presents a
rich dynamics since a resonance structure is visible at center of
mass energy $E\approx3$ MeV, as already shown in
Fig.~(\ref{Diff_234N}).

Up to now only two of the techniques used in bound state calculations
(FY and HH) as well as methods based on the resonating group model
(RGM)~\cite{PHH01} and  on the solution of Alt, Grassberger and
Sandhas (AGS) equations~\cite{AF_99,GSAGS_67}
have been employed to attack this problem
with realistic Hamiltonians (for a review of earlier results, see
Ref.~\cite{TWH92}). Some previous FY calculations in configuration
space suggested that realistic NN interactions could fail in
reproducing  the experimental total and differential cross
sections~\cite{CCG_98,CCGF_98,AF_99}. These calculations were
however limited to a relatively small number of partial waves in
the expansion of FY amplitudes, thereby such lack of convergence
could be advocated to explain the failure. Using AGS equations,
the author of Ref.~\cite{AF_99} was able to substantially increase
the number of partial wave amplitudes and obtain a fairly good
agreement for the total cross section at the resonance peak.
However, these calculations were based on a rank-one separable
expansion of the 2-body t-matrix, which also raises some doubts on
the reliability of the results. The calculations performed by
expanding the scattering wave functions on the
correlated Hyperspherical Harmonic (CHH) basis were able
to obtain a reasonable estimate of the various phase-shift and
mixing-angle parameters~\cite{Viv98,Viv01}. However, the problem
of convergence could not be solved completely due to numerical
difficulties when adding more CHH components. For this reason, a
new expansion on the (uncorrelated) HH basis is considered here.

The results obtained using these techniques were so far at rather
large variance between each other. Clearly, this situation should
be clarified before questioning the ability of present NN + 3N
force models to describe the experimental data beyond the binding
energy of $^4$He, for which they have proven to be rather
successful~\cite{Wea00}. This is the purpose of the present paper
in which we compare low energy $n-\nuc{3}{H}$ scattering results
obtained by three different groups, using independent methods to
solve the four-body equations. They concern AGS
calculations~\cite{AF_99} based on rank-one expansion of the NN
t-matrix, variational solutions of the Schr\"odinger equation
using HH method~\cite{VKR_04} and the solutions of FY
equations~\cite{Fred_97,LC_FB17_03} with a substantial improvement
in the number of partial waves~\cite{LC_Trento_03};
AV18~\cite{AV18} and NIJM-II~\cite{NIJ_93} NN interaction models
are used in present calculations.

We have considered the center of mass energies
$E=0.40,0.75,1.50,2.625,3.0$ MeV.
They constitute a good set of data for interpolating other values of
the phase shifts and allow to determine the low energy parameters
(scattering length and effective range).
In addition, values $E=0.75,1.50, 2.625$ MeV correspond to the measured differential
cross sections.

To describe scattering solutions, we will expand the different
$J^{\pi}$ $n-\nuc{3}{H}$ states in terms of the  asymptotic hamiltonian
channels $|L,S;J^{\pi}>$,
where $L$ denotes the $n-\nuc{3}{H}$ relative orbital angular momentum and
$\vec{S}=\vec{s}_n+\vec{s}_t$ the total spin.
The following $J^{\pi}$ states will be considered:
\begin{eqnarray}
|0^{+}\rangle=&       &|0,0;0^{+}\rangle                 \cr
|1^{+}\rangle=&c_{1^+}&|0,1;1^{+}\rangle + d_{1^+}|2,1;1^{+}\rangle \cr
|0^{-}\rangle=&       &|1,0;0^{-}\rangle                  \cr
|1^{-}\rangle=&c_{1^-}&|1,1;1^{-}\rangle + d_{1^-}|1,0;1^{-}\rangle \cr
|2^{-}\rangle=&c_{2^-}&|1,1;2^{-}\rangle + d_{2^-}|3,1;2^{-}\rangle
|\label{ntstates}
\end{eqnarray}
Coefficients $c$ and $d$ are fixed by the dynamics and make
the solutions eigenstates of the S-matrix.

As it will be demonstrated later, these states are the only relevant ones for the low energy scattering considered here.
Moreover,  the coupling $L\leftrightarrow L+2$ is very weak and is neglected for $J^{\pi}=2^{-}$ state.

\section{Methods}

We briefly describe the methods used for solving the 4N problem.
Two of them (FY and HH) work in configuration space while AGS in
momentum space.

\subsection{AGS equations}

The starting point involves the AGS equations~\cite{GSAGS_67} for
the transition operators involving all (2N)+(2N) and N+(3N)
channels. For local NN potentials such equations are three-vector
variable integral equations which after partial wave decomposition
reduce to a set of coupled equations in three continuous scalar
variables. Since scattering calculations require a great number of
channels for convergence, we follow an approach based on the
separable representation of subsystem amplitudes in order to
reduce the equations to two or one continuous variable. The
integral equations we use are the same as in Ref.~\cite{AF_89.1}
and result from the modified AGS equations~\cite{HS_81} after
one has: (a) represented the original NN t-matrix by an operator
of rank one; (b) represented the resulting 3N t-matrix by a
finite rank operator and taken as many terms as needed for
convergence. Since in the modified AGS equations the 2N+2N
subamplitudes are expressed in terms of a convolution integral
involving two non-interacting pair-propagators, as first
proposed in Ref.~\cite{FS_76}, the sole approximation in this
approach involves a rank one representation of the 2N t-matrix
which may be obtained from the well-known method of Ernst,
Shakin, and Thaler (EST)~\cite{EST_73}. The multi-term
representation of the 3N t-matrix is done using the Energy
Dependent Pole Expansion (EDPE) method developed in
Ref.~\cite{SMcGF_79}. This latter representation for the 3N
t-matrix is well under control since one may check the
convergence rate of 4N observables for increasing rank in the
3N t-matrix.

This method was first used in Ref.~\cite{AF_89.1} to calculate
the binding energy of $^4$He and later confirmed to be accurate by the
exact work of Kamada and Gl\"ockle~\cite{KG92}. More
recently~\cite{CCGF_98} the results of our calculations for
$n-\nuc{3}{H}$ elastic scattering were shown to agree with the exact results of the
Grenoble group~\cite{CCG_99}, for both Malfliet-Tjon (MT) and AV14
potentials taken in 2N partial waves with total angular momentum j
$\leq1^+$ ($^1{\rm S}_0, ^3{\rm S}_1-^3{\rm D}_1$). In the present
calculations we extend the work in Ref.~\cite{AF_99} by increasing the
number of 2N partial waves to
$j \leq 2$. Therefore we add $^3{\rm F}_2$, $^1{\rm D}_2$ and
$^3{\rm D}_2$ to the partial waves $^1{\rm S}_0,
^3{\rm S}_1-^3{\rm D}_1$, $^1{\rm P}_1$, $^3{\rm P}_0$, $^3{\rm
P}_1$ and $^3{\rm P}_2$ already included in Ref.~\cite{AF_99}.
Moreover we limit the particle-pair orbital angular momentum $\ell_y
\leq 2$ and include all
$J_3$ and
$\ell_z \leq 2$ for a given total four-body angular momentum $J$; $J_3$
is total angular momentum of any given three-body subsystem and $\ell_z$
is the orbital angular momentum of that subsystem relative to the fourth
particle. The results we find here are consistent with those obtained
in Ref. ~\cite{AF_99} and converged relative to the number of terms in
the finite rank expansion of the underlying 3N t-matrix. The number of
terms in the expansion range from four to six depending on
$J_3$.

\subsection{Faddeev-Yakubovsky method}

In the case of four identical fermions, interacting via a pair-wise
potential $V$,
the FY equations result into a set of two integrodifferential equations,
coupling two FY components, namely K and H:
\begin{eqnarray}
\left( E-H_{0}-V\right) K &=&V(P^{+}+P^{-})\left[
(1+Q)K+H\right] \\
\left( E-H_{0}-V\right) H &=&V\tilde{P}\left[
(1+Q)K+H\right]  \label{FYE}
\end{eqnarray}%
with $P^{+}$, $P^{-}$, $\tilde{P}$ and $Q$ being particle permutation
operators:
\begin{eqnarray}
\begin{tabular}{lll}
$P^{+}=(P^{-})^{-}=P_{23}P_{12};$ & $Q=\varepsilon P_{34}$; & $\tilde{P}%
=P_{13}P_{24}=P_{24}P_{13}$,%
\end{tabular}%
\end{eqnarray}%
and $\varepsilon$ is a Pauli factor for exchange of two identical
particles, which in case of fermions is -1. The wavefunction is
given by:
\begin{eqnarray}
\Psi =\left[ 1+(1+P^{+}+P^{-})Q\right]
(1+P^{+}+P^{-})K+(1+P^{+}+P^{-})(1+\tilde{P})H\label{psi22}
\end{eqnarray}%

Each FY component F=(K,H) is considered as a function, described in its proper set of
Jacobi coordinates
$\vec{x},\vec{y},\vec{z}$, defined respectively by
\begin{equation}
  \begin{array}{ccc}
  \vec{x}_K&=& \vec{r}_2-\vec{r}_1 \cr
  \vec{y}_K&=& \sqrt{4 \over3}\left(\vec{r}_3-
  {\vec{r}_1+\vec{r}_2\over2}\right)\cr
  \vec{z}_K&=& \sqrt{3 \over2}\left(\vec{r}_4-
  {\vec{r}_1+\vec{r}_2+\vec{r}_3\over3}\right)
  \end{array}\qquad
  \begin{array}{ccc}
  \vec{x}_H&=& \vec{r}_2-\vec{r}_1 \cr
  \vec{y}_H&=& \vec{r}_4-\vec{r}_3 \cr
  \vec{z}_H&=& \sqrt{2}\left({\vec{r}_3+\vec{r}_4\over2}-
  {\vec{r}_1+\vec{r}_2\over2}\right)
  \end{array}
\end{equation}
and expanded in angular variables for each coordinate according to
\begin{equation}\label{KPW}
  \langle\vec{x}\vec{y}\vec{z}|F\rangle=
  \sum_{\alpha} \; {F_{\alpha}(xyz)\over xyz} \;Y_{\alpha}
  (\hat{x},\hat{y},\hat{z})  .
\end{equation}
The quantities $F_{\alpha}$ are called regularized FY amplitudes and
$Y_{\alpha}$ are
tripolar harmonics, containing spin, isospin and angular momentum
variables.  The label $\alpha$ holds for the set of 10 intermediate
quantum numbers describing a $J^{\pi},T=1$ state . They are defined in a
j-j coupling scheme as
\begin{eqnarray*}
  K&\equiv&\left\{\left[(t_1t_2)_{\tau_x}t_3\right]_{T_3}t_4
  \right\}_{T=1}
  \otimes\left\{ \left[ \left( l_x (s_1 s_2)_{\sigma_x} \right)_{j_x}
  (l_y s_3)_{j_y}
  \right]_{J_3} (l_z s_4)_{j_z} \right\}_J\cr
  H&\equiv&\left[ (t_1 t_2)_{\tau_x} (t_3 t_4)_{\tau_y}
  \right]_{T=1}\otimes
  \left\{ \left[ \left( l_x (s_1 s_2)_{\sigma_x} \right)_{j_x}
  \left( l_y (s_3 s_4)_{\sigma_y} \right)_{j_y} \right]_{j_{xy}} l_z
  \right\}_J
\end{eqnarray*}
where $s_i$ and $t_i$ are the spin and the isospin of the
individual particles and $T,J$ the isospin and total angular
momentum of the four-body system. Each of the $N_c=N_K+N_H$
amplitudes in the expansion (\ref{KPW}) is labelled by 12 quantum
numbers, which are further conditioned by the antisymmetry
properties $(-)^{\sigma_x+\tau_x+l_x}=\varepsilon$ for K and
$(-)^{\sigma_x+\tau_x+l_x}=(-)^{\sigma_y+\tau_y+l_y}=\varepsilon$
for H.

The boundary conditions for the 1+3 scattering problem are implemented
by imposing at large enough value of $z$ the Dirichlet-type condition
\begin{eqnarray*}
  K(x,y,z) &=&  t(x,y) \\
  H(x,y,z) &=&  0
\end{eqnarray*}
$t(x,y)$ being the triton Faddeev component with quantum numbers
\(
  \left[\left(l_x(s_1 s_2)_{\sigma_x}\right)_{j_x}(l_y
  s_3)_{j_y}\right]_{J_3}
\).
They ensure a solution which, e.g. for a relative $n-\nuc{3}{H}$ S-wave, behaves
asymptotically like
\begin{eqnarray*}
  K(x,y,z) &\sim&  t(x,y)\sin{(qz+\delta)}
\end{eqnarray*}
where $\delta$ is the $n-\nuc{3}{H}$ phase shift and $q$, the conjugate momentum
of the $z$-Jacobi coordinate in K-amplitudes,
is related to the center of mass $n-\nuc{3}{H}$ kinetic energy $E_{\mathrm{cm}}$
and the physical momentum $k$ by
\begin{equation}\label{eq:k}
 E_{\mathrm{cm}}={3\over4}E_{\mathrm{lab}}={\hbar^2\over
  m}q^2={2\over3}{\hbar^2\over m}k^2 \ .
\end{equation}
The convergence is reached in respect to expansion (\ref{KPW}).

\bigskip
FY calculations have been performed in the $j-j$ coupling scheme.
The following truncations in the partial wave expansion of amplitudes were used:
{\it (i)} V$_{NN}$ waves limited to $l_x\le3$, always including
tensor-coupled partners, i.e. involving the set
$^1S_0,^3SD_1$, $^1P_1,^3P_0,^3PF_2,^3P_1$, $^1D_2,^3DG_3,^3D_2$,
$^1F_3,^3FH_4,^3F_3$ and {\it (ii)} $l_x+l_y+l_z\le10$.

The convergence was studied as a function of
$j_{yz}$=max($j_y$,$j_z$) for K-like components and
$j_{yz}$=max($j_y$,$l_z)$ for H-like, starting with $j_{yz}=1$. In
the second column of the Table \ref{Conv_delta_FY} we present the
values of the relevant phase shifts  for AV18 model at the peak
energy $E_{cm}$=3.0  MeV. The couplings $L\leftrightarrow L+2$ in
the asymptote of the $1^+$ ($L=0\leftrightarrow L=2$) and $2^-$
($L=1\leftrightarrow L=3$) states are found to be very small (see
next section) and have been omitted here. For the $1^-$  state, we
give the eigenphaseshifts - $\delta_1(1^-),\delta_2(1^-)$  -- and
the mixing parameter $\epsilon$. One can remark that results
displayed on Table \ref{Conv_delta_FY} converge pretty well as a
function of $j_{yz}$. For the most of the phaseshifts three-digit
accuracy is reached already with  $j_{yz}=2$. The most unstable
were found J$^{\pi}=2^-$ phaseshift and J$^{\pi}=1^-$ mixing
parameter, which were requiring inclusion of $j_{yz}=3$ amplitudes
to converge at a one percent level.

\subsection{Hyperspherical harmonic method}

In the HH method, the wave function is written as the sum of an
asymptotic part, which form is known (except for the parameter of
the $S$-matrix), and an internal part, which is expanded in HH
functions. Explicitly, the wave function $\Psi_{1+3}^{LSJJ_z}$
describing a $n-\nuc{3}{H}$ scattering state with incoming orbital
angular momentum $L$ and channel spin $S$ ($S=0, 1$) coupled to
total angular $JJ_z$, is expressed as
\begin{equation}
  \Psi_{1+3}^{LSJJ_z}=\Psi_C^{JJ_z\pi}+\Psi_A^{LSJJ_z} \ ,
  \qquad \pi=(-)^L\ ,
  \label{eq:psica}
\end{equation}
where $\Psi_C^{JJ_x\pi}$ vanishes in the limit of large
intercluster separations, and hence describes the system in the
region where the particles are close to each other and their
mutual interactions are strong. On the other hand,
$\Psi_A^{LSJJ_z}$ describes the relative motion of the two
clusters in the asymptotic regions, where the $n-\nuc{3}{H}$ interaction
is negligible and can be decomposed as a
linear combination of the following functions
\begin{equation}
  \Omega_{LSJJ_z}^\pm=\frac{1}{\sqrt{4}}
  \sum_{i=1}^4
  \Bigl[ [ s_i \otimes \phi_3(jlp) ]_{S} \otimes
  Y_{L}(\hat{\bf r}_i) \Bigr ]_{JJ_z}
  \Bigl ( -f_R(r_i) y_{L}(kr_i)\pm {\rm i} j_L(kr_i) \Bigr ) \ ,
  \label{eq:psiom}
\end{equation}
where $r_i$ is the distance between the neutron (particle $i$) and
$\nuc{3}{H}$ (particles $jlp$), $k$ is the magnitude of the
relative momentum between the two clusters defined in
Eq.~(\ref{eq:k}), $\phi_3$ is the $\nuc{3}{H}$ wave function,
which has been obtained by solving the corresponding three-body
problem using the pair-correlated HH expansion~\cite{KRV94} and
$j_L$ and $y_L$ are the spherical Bessel functions of the first
and second kind, respectively.
 The function $f_R(r_i)$ has been
introduced to regularize  $y_L(kr_i)$ at small $r_i$, and
$f_R(r_i) \rightarrow 1$ as $r_i$ is large, thus not affecting the
asymptotic behavior of $\Psi_{1+3}^{LSJJ_z}$. Note that for large
values of $kr_i$,
\begin{equation}
  -f_R(r_i) y_{L}(kr_i)\pm {\rm i} j_L(kr_i) \rightarrow
   { \cos(k r_i-L\pi/2) \pm {\rm i} \sin(k r_i-L\pi/2) \over k r_i}
  ={ \exp\Bigl[\pm {\rm i} (k r_i-L\pi/2) \Bigr] \over kr_i}\ .
\end{equation}
Therefore, $\Omega_{LSJJ_z}^+$  ($\Omega_{LSJJ_z}^-$) describes
in the asymptotic regions an outgoing (ingoing) $n-\nuc{3}{H}$ relative
motion. Finally,
\begin{equation}
  \Psi_A^{LSJJ_z}= \sum_{L^\prime S^\prime}
   \bigg[\delta_{L L^\prime} \delta_{S S^\prime} \Omega_{LSJJ_z}^-
   - {\cal S}^J_{LS,L^\prime S^\prime}(k)
     \Omega_{LSJJ_z}^+ \bigg] \ ,
  \label{eq:psia}
\end{equation}
where the parameters ${\cal S}^J_{LS,L^\prime S^\prime}(k)$ are the
$S$-matrix elements which determine phase-shifts and (for coupled
channels) mixing angles at the energy
$(2/3) k^2/m$ .  Of course, the sum over $L^\prime$ and
$S^\prime$ is over all values compatible with a given $J$ and parity.

The \lq\lq core\rq\rq wave function $\Psi^{JJ_z\pi}_C$ has been here
expanded in the HH basis as for the bound-state wave
function~\cite{VKR_04},
\begin{equation}\label{eq:chh2}
     \Psi_C^{JJ_z\pi}=  \sum_{\alpha=1}^{N_c}\;
      \sum_{n_2,n_3=0}^{n_2+n_3\le N(\alpha)}
      u_{\alpha n_2 n_3}(\rho)
      {\cal Y}_{\alpha n_2 n_3}(\vec{x}_K,\vec{y}_K,\vec{z}_K)
      \ ,
\end{equation}
where $\rho$ is the hyperradius, $\rho=\sqrt{x_K^2+y_K^2
+z_K^2}$.  The known functions ${\cal Y}_{\alpha
  n_2 n_3}(\vec{x}_K,\vec{y}_K,\vec{z}_K)$ are given as
the antisymmetrized product of spin-isospin states and HH
functions~\cite{F83}, the latter being given by the product of
tripolar spherical harmonics and two Jacobi polynomials of indexes
$n_2$ and $n_3$ given, respectively, in terms of ``hyperangles''
defined by
\begin{equation}
    \cos\varphi_{2} = { y_K \over \sqrt{z_K^2+y_K^2}}\ ,
    \qquad
    \cos\varphi_{3} =
    { x_K \over \rho} \ .
     \label{eq:phi}
\end{equation}
The channels index $\alpha$ denotes collectively the orbital
angular momenta, spin and isospins quantum numbers. For a given
channel $\alpha$, the HH states with indexes $n_2$ and $n_3$
chosen in the range $0\le n_2+n_3\le N(\alpha)$ are included in
the expansion. The values of the number of channels $N_c$ and the
various $N(\alpha)$ are increased until the desired degree of
convergence in the quantity of interest is obtained (for the
choice of which channels to include see the discussion below).

The functions $u_{\alpha n_2 n_3}(\rho)$ occurring in the expansion of
$\Psi^{JJ_z\pi}_C$ and  the matrix elements ${\cal S}^J_{LS,L^\prime
  S^\prime}(k)$  are determined by making the functional
\begin{equation}
   [\overline{\cal S}^J_{LS,L^\prime S^\prime}(k)]=
    {\cal S}^J_{LS,L^\prime S^\prime}(k)
     -\frac{m}{\sqrt{6}{\rm i} }
        \langle\Psi^{L^\prime S^\prime JJ_z }_{1+3} |
         H-E_3-T_{\mathrm{cm}} |
        \Psi^{LSJJ_z}_{1+3}\rangle
\label{eq:kohn}
\end{equation}
stationary with respect to variations in the ${\cal
S}^J_{LS,L^\prime S^\prime}$ and $u_{\alpha n_2 n_3}$ (Kohn variational
principle).  Here $E_3$ is the $\nuc{3}{H}$ ground-state energy. By
applying this principle, a set of second order differential
equations for the functions $u_{\alpha n_2 n_3}(\rho)$ are obtained. By
replacing the derivatives with finite differences, a linear system
is obtained,which have been solved using the Lanczos algorithm.
The procedure is very similar to that outlined in the appendix of
Ref.~\cite{KMRV96} and it will not be repeated here.
A large number of equation can be quite easily be solved.

Let us now discuss briefly the choice of the HH states to be
included in the expansion. As for the $\alpha$-particle bound
state~\cite{VKR_04}, the ``brute force'' inclusion of HH states
is not possible  and one has to select the appropriate
subset of functions. We have found convenient~\cite{VKR_04}
to separate the HH functions in classes depending on the value
of ${\cal L}=\ell_x+\ell_y+\ell_z$: channels with low values as
possible of ${\cal L}$ have to be included in the expansion first.
For each group of channels of a given ${\cal L}$, the corresponding
values of $N({\alpha})$ are increased until convergence is
reached. Then, channels with a larger value of ${\cal L}$ are
included in the expansion and so on. The advantage of this
procedure is that the classes of channels with large values
of ${\cal L}$ have small contributions. Consequently, the corresponding
values of $N({\alpha})$ can be also taken smaller and smaller.

An example of convergence is presented in Table~\ref{Conv_delta_FY},
in the last three columns. There, the phase shifts at $E_{cm}=3$ MeV
obtained by including in the HH expansion the channels with
increasing values of ${\cal L}$ have been reported.
For S-wave phase shifts, the inclusion of the ${\cal L}=0$ channels
is sufficient to have already rather good estimates. The contribution
of the ${\cal L}\ge 4$ channels  is found to be negligible.
For the P-wave phase shifts, the channels with ${\cal L}=1$ give
the dominant contribution at low energies ($E_{c.m.}\le 1.5$ MeV),
but in the peak region they are clearly insufficient to obtain
good estimates. Including the channels with ${\cal L}=3$, the
HH predictions already differ less of 4\%
with respect to the final FY results.
There exist a very large number
of ${\cal L}=5$ channels and we have limited ourselves
to the inclusion of those (usually) most important, namely, those having
$\ell_x=\ell_y=2$ and $\ell_z=1$, plus permutations
(this is sufficient to double the number of channels).
However, the changes of the P-wave phase shifts
are rather tiny. Therefore, we expect that the contribution
of the disregarded channels ${\cal L}=5$ (and those with ${\cal L}>5$)
to be practically negligible. This issue is currently being
investigated~\cite{VKRip}.

It must be noticed that in the present HH calculation only states
constructed in terms of the Jacobi vectors of the set K
have been considered (hereafter referred to set-K HH functions).
Obviously, the HH functions of a given set of
Jacobi vectors form a complete basis by themselves.
However, the HH functions constructed with the set H of Jacobi vectors
(set-H HH functions) should be more suitable for taking into account
contributions from the $\{2+2\}$ cluster structures.
It is rather obvious that the inclusion of HH functions of both sets
should speed up the convergence in constructing the state of the
system. In the present calculation, however, we have included
only the set-K HH functions since we were able to include a number of
states sufficient to reach, in the studied cases,
a satisfactory degree of convergence.

\begin{table}[h!]
\begin{center}
\caption{Convergence of $n-\nuc{3}{H}$ phase shifts and mixing parameter
for AV18 potential at E$_{cm}$=3 MeV. The coupling $L\to L+2$
in the asymptotics for the $J^\pi=1^+$ and $2^-$ states
have been neglected. See the text for explanations.}
\label{Conv_delta_FY}
\begin{tabular}{l||cccc|ccc|}
\hline
   & \multicolumn{4}{c|}{FY} & \multicolumn{3}{c|}{HH} \\
\hline
J$^{\pi}$ &$j_{yz}\le1$&$j_{yz}\le2$& $j_{yz}\le3$& $j_{yz}\le4$ &
   ${\cal L}=0$ & ${\cal L}=2$ & ${\cal L}=4$ \\
\hline
0$^+$     &   -70.11   & -69.98     & -70.00      & -70.00 &
           -70.1 & -69.8 & -69.8 \\ \hline
1$^+_1$   &   -63.61   & -61.94     & -61.97      &        &
     -62.6 & -62.2 & -62.2 \\ \hline\hline
J$^{\pi}$ &$j_{yz}\le1$&$j_{yz}\le2$& $j_{yz}\le3$& $j_{yz}\le4$ &
${\cal L}=1$ & ${\cal L}=3$ & ${\cal L}=5$ \\
\hline
0$^-$     &    26.28   &  23.86     &  23.60      &  23.58  &
   16.9 & 23.0 & 23.4 \\ \hline
1$^-_1$   &    25.32   &  22.40     &  22.27      &         &
   20.7 & 22.0 & 22.2 \\
1$^-_2$   &    18.00   &   40.70    &  40.88      &         &
   31.5 & 41.5 & 42.0 \\
$\epsilon$&     9.04   &    -43.52  &   -44.39    &         &
  -40.9  & -44.9  & -45.3\\\hline
2$^-$     &     -      &     43.82  &   45.41     &   45.44    &
 27.9 & 44.0 & 44.5  \\
\hline
\end{tabular}
\end{center}
\end{table}

\newpage
\section{Results}

The $n-\nuc{3}{H}$ scattering lengths have already
been presented in Ref. \cite{Viv98,RL_Thesis,AF_99}.
For completeness, in Table \ref{Tab_sc_length} we summarize the latest
obtained values with the improved precision.

In Table \ref{Tab_depha_SP_AV18} we present phase shifts,
mixing parameters and the corresponding total cross sections
for the most relevant
partial waves  at some selected energies.
The coupling
$L\leftrightarrow L+2$ between the different asymptotics in the
$J^{\pi}=1^{+},2^{-}$ states turned out to be very small. In AGS
and FY calculations they still have been taken into account for
$J^{\pi}=1^{+}$ but neglected for $J^{\pi}=2^{-}$ state. In fact,
even at E$_{cm}=3$ MeV, the largest energy we have considered,
$L=2$ phaseshift in $J^{\pi}=1^{+}$ is only $\sim1^{\circ}$ and
contributes by only 0.01\% in total cross section. Contribution of
$L=3$ phaseshifts, which were neglected in $J^{\pi}=2^{-}$
calculations, is at least by one more order of magnitude smaller.

\begin{table}[h!]
\begin{center}
\caption{Singlet a$_1$(J$^\pi$=0$^+$) and triplet a$_3$(J$^\pi$=1$^+$) scattering lengths calculated with Nijm-II
and AV18 NN interactions.}\label{Tab_sc_length}

\begin{tabular}{l||ccc|l}\hline
Pot. & a$_{1}$ (fm)  & a$_{3}$ (fm) & $\sigma(0)$ (b)   &Method \\
\hline\hline
             &     4.09        &  3.70     &  1.82      &  AGS   \\
  AV18       &     4.25        &  3.74     &  1.88      &  FY    \\
             &     4.28        &  3.73     &  1.89      &  HH    \\
\hline\hline
  Nijm-II    &     4.09        &  3.70     &  1.82      &  AGS   \\
             &     4.25        &  3.74     &  1.88      &  FY    \\
 \hline\hline
\end{tabular}
\end{center}
\end{table}

One can see that positive parity states are under full control by
the different methods. Agreement between HH and FY
results is perfect for positive parity phaseshifts (in the most of
cases all three significant digits coincide, whereas the largest
discrepancy does not exceed 0.3\%).
AGS results also agree by a few percent.

The situation is more delicate for negative parity states, where
resonances are present and various techniques suffer from slower
convergence. Nevertheless FY and HH results stay in close agreement to
better than few percent for phaseshifts, though it could be perfect if
one or more convergence step is effected (see Table  \ref{Conv_delta_FY}).
Still these small, convergence related, correlations do not shelter the
excellent agreement in total cross sections. On the other hand
disagreement with AGS results becomes important, reaching 30\% for the
phaseshifts in $J^{\pi}=0^{-}$ state. All the negative parity
phases are strongly overestimated by this method, thus resulting a
larger total cross sections in the resonance region.

We believe that the underlining reason for the disagreement with AGS
results is due only to the rank-one separable expansion for 2-body
t-matrix used in this calculations. This lowest order
expansion may be enough to describe systems dominated by spherical
symmetric wave functions (i.e. positive parity states) but fails for more
complicated structures, not being able to account for strong
compensations present in NN P-waves
\cite{RL_Thesis}, which turn out to be sizeable in negative parity
states \cite{CarbLaz_PRC04}.

\begin{table}[h!]
\begin{center}
\caption{AV18 $n-\nuc{3}{H}$ S- and P-wave phase-shifts (degrees)
as a function of the c.m. kinetic energy $E_{\mathrm cm}$ (MeV).
Numbers in brackets are the mixing parameters in degrees for
corresponding scattering states. }\label{Tab_depha_SP_AV18}

\begin{tabular}{l||l| lll |c|ccc|c|c|l}\hline
E$_{cm}$&0$^+$ & 1$^+_1$ & 1$^+_2$&     & 0$^-$&1$^-_1$&1$^-_2$&
&2$^-$ &$\sigma$ (b)&\\\hline
0.40    &-27.8 &-25.0   &-0.018 &      & 2.24 & 2.87  & 3.89  &
& 3.56 & 1.73  & AGS\\
        &-28.8 &-25.1   &-0.0165 &(0.199) & 1.81 & 2.66  & 3.56
        &(-39.6)
& 3.38 & 1.75  & FY \\
        &-28.8 &-25.1   &        &      & 1.78 & 2.70  & 3.76
        &(-44.2)
& 3.32 & 1.76  & HH \\ \hline
0.75    &-37.3 &-33.6   &-0.072  &      & 5.59 &  6.69 & 9.95  &
        & 9.15 & 1.79  & AGS\\
        &-38.7 &-33.8   &-0.059  &(0.358) & 4.41 &  6.11 & 8.85  &
        (-41.3)
        & 8.57 & 1.78  & FY \\
        &-38.7 &-33.8   &        &      & 4.36 &  6.08 & 9.32  &
        (-45.4)
        & 8.53 & 1.79   & HH\\ \hline
1.50    &-50.7 &-46.1   &-0.274  &      & 14.5 &  15.1 & 25.9  &
        & 24.3 & 2.22  & AGS\\
         &-52.8 &-46.2   &-0.330  &(0.670) & 11.5 &  13.4 & 21.9  &
         (-42.9)
        & 22.5 & 2.06  & FY \\
    &-52.6 &-46.3   &        &      & 10.8 &  13.3 & 23.0  &
    (-45.6)
        & 22.1 & 2.06  & HH\\ \hline
2.625   &-63.7 &-58.3   &-0.649  &      & 27.5 &  24.7 & 44.8  &
        & 44.2 & 2.51  & AGS\\
          &-66.5 &-58.5   &-0.850  &(1.08)  & 20.9 &  20.7 & 37.3
          &(-43.5)
        & 41.0 & 2.24  & FY \\
        &-66.3 &-58.7   &        &      & 20.6 &  20.5 & 38.6  &
        (-45.5)
        & 40.1 & 2.24     & HH\\ \hline

3.0     &-67.4 &-61.9   &-0.764  &      & 30.9 &  26.8 & 48.5  &
        & 48.5 & 2.48  & AGS\\
         &-70.0 &-62.0   &-1.180  &(1.21)  & 23.9 &  22.4 & 40.7
         &(-43.5)
        & 45.4 & 2.21  & FY \\
        &-69.8 &-62.2   &        &      & 23.4 &  22.2 & 42.0  &
        (-45.3)
        & 44.5 & 2.21      & HH  \\ \hline
\end{tabular}
\end{center}
\end{table}

\begin{figure}[h!]
\begin{center}\mbox{\epsfxsize=12.cm\epsffile{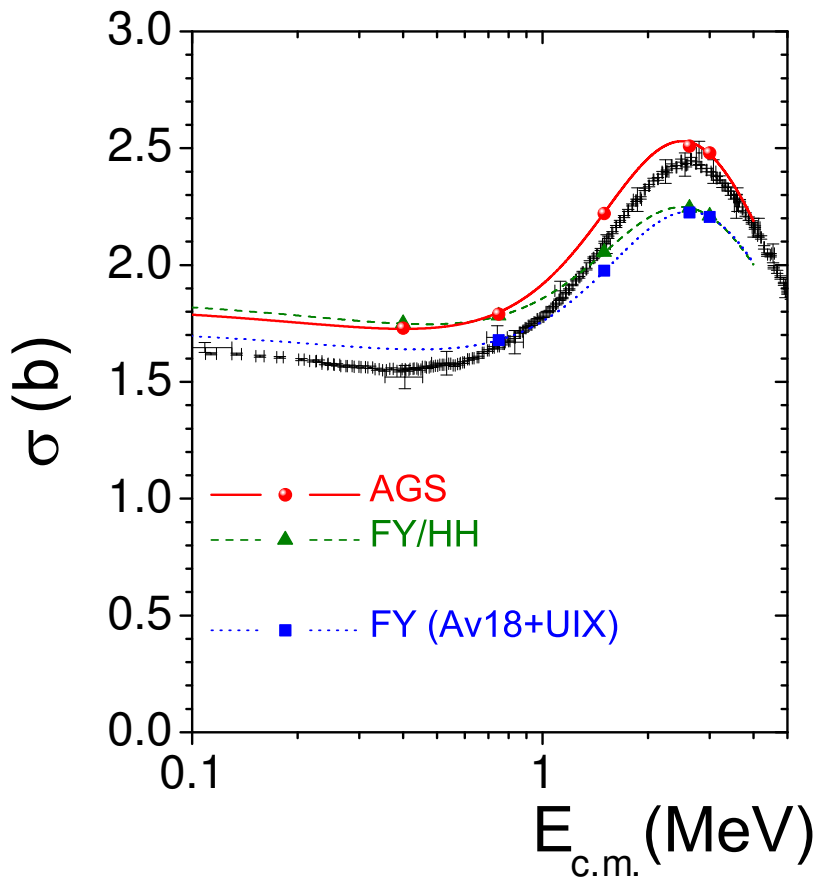}}\end{center}
\vspace{-1.0cm}
\caption{Comparison between experimental and
theoretical n-$^3$H total cross section calculated using different
methods and with AV18 potential. Dotted curve presents AV18+UIX
model results of Ref.~\cite{CarbLaz_PRC04}.}\label{sig_nt}
\end{figure}

The  total cross sections are plotted in Fig. \ref{sig_nt} and
compared to the experimental values taken from Ref. \cite{PBS_80}.
The results of HH and FY (dashed line) are in agreement for three
significant digits and thus are not distinguishable by eye. The
AGS cross sections (solid line) are slightly smaller at very low
energies, whereas they provide considerably larger cross sections
in the resonance region.
The disagreement of theoretical results with experimental data at
zero energy can be attributed to the $^3$H underbinding by local
NN interaction models. It can be improved by including
a three nucleon force (3NF), i.e.
the Urbana version IX (UIX), and thus reproducing the
3N binding energies
\cite{CCG_99,Viv98,LC_FB17_03,LC_Trento_03}. In fact, HH and FY results
for AV18+UIX model agree on $\sigma(0)=1.73$ b total cross section
value, which is consistent
with experimental one $\sigma(0)=1.70\pm0.03$ b  of Ref. \cite{PBS_80}.
Three nucleon forces, and UIX in
particular, have however very small effect in the resonance
region as shown in Fig \ref{sig_nt}; it has even the tendency to
diminish the cross section due to triton rescaling
\cite{CCG_99,LC_FB17_03,LC_Trento_03}. The very weak
sensitivity of total cross sections to ``standard'' 3NF,
may indicate that this disagreement could have its origin
in the NN forces themselves and their P-waves
in particular~\cite{CarbLaz_PRC04}.

Let us remember that in $N-d$ scattering there exist
other large discrepancies between theory and experiment,
the most known example is the so called $A_y$
puzzle~\cite{KH86,WGC88,KRTV96}.
To explain these discrepancies, speculations about either the
deficiency of the NN potentials in ${}^3P_j$  waves~\cite{WG91},
or the presence of exotic 3N force
terms not contemplated so far~\cite{K99,CS01}, have been advanced.
Whether such discrepancies in 3N and 4N systems
originate from the same deficiency in the
nuclear Hamiltonian is still under debate.

Some doubts could be casted on the reliability of the experimental
data in the resonance region, due to existence of a single
accurate data set~\cite{PBS_80}. However, due to the near agreement
with calculations at low energies, its hardly believable that
experimental points could contain a normalization error near the
resonance peak. A normalization error can also be excluded by
analyzing differential cross sections~\cite{RL_Thesis}: they indicate
that theoretical results agree with experimental ones at
$\theta_{cm}=90$ deg, where positive parity phaseshifts dominate, while
diverge at forward and backward directions, where negative parity
phaseshifts become important. Therefore the experimental data of
ref.~\cite{PBS_80} seems to be accurate and coherent. In addition,
at near-resonance energies, this data seems to be also in agreement with
older and less accurate measurements~\cite{LACG_59}. Nevertheless, in
view of the very few experimental studies of n-$\nuc{3}{H}$ it
would be very interesting to have an independent confirmation
of the data.

\bigskip
A similar analysis has been performed with FY and AGS methods
using Nijm-II model. The obtained results are summarized in Table
\ref{Tab_depha_SP_NijmII}. Note that a particular choice of the
local NN-interaction model has no qualitative impact. Thus Nijm-II
predictions are very close to the AV18 ones and lead to the same
discrepancies.

\begin{table}[h!]
\begin{center}
\caption{Nijm-II $n-\nuc{3}{H}$ S- and P-wave phase-shifts
(degrees) as a function of the center of mass kinetic energy
$E_{\mathrm cm}$ (MeV). Numbers in brackets are the mixing
parameters in degrees for corresponding scattering
states.}\label{Tab_depha_SP_NijmII}

\begin{tabular}{l||c|ccc|c|ccc|c|c|l}\hline
E$_{cm}$& 0$^+$ &1$^+_1$&1$^+_2$&      & 0$^-$ &1$^-_1$ &1$^-_2$&
&  2$^-$ &$\sigma$ (b)&\\\hline
0.40    & -27.81&-25.03 &-0.0183&      & 2.28  &  2.84  & 3.89  &
& 3.47   &  1.73 & AGS\\
        &-28.66 &-25.03 &-0.0167&(0.196)& 1.751 &  2.633 & 3.537
&(-39.44) &  3.349 &  1.74 & FY \\ \hline
0.75    &-37.27 &-33.68 &-0.072 &      & 5.67  &  6.63  & 9.94  &
        &  8.92  &  1.79 & AGS\\
        &-38.57 &-33.73 &-0.696 &(0.354) & 4.252 &  6.040 & 8.763
        &(-41.03) &  8.505 &  1.77 & FY \\ \hline
1.50    &-50.68 &-46.11 &-0.271 &      & 14.66 & 14.98  & 25.83 &
        & 23.87  &  2.21 & AGS\\
        &-52.48 &-46.22 &-0.334 &(0.661) & 10.74 &  13.20 & 21.72
        &(-42.69) &  22.37 &  2.04 & FY \\ \hline
2.625   &-63.75 &-58.38 &-0.639 &      & 27.77 & 24.51  & 44.66 &
        & 43.69  &  2.51 & AGS\\
        &-66.49 &-58.51 &-0.813 &(1.067) & 20.22 &  20.44 & 36.96
        &(-43.31) &  40.82 &  2.23 & FY \\ \hline
3.0     &-67.18 &-61.81 &-0.790 &      & 32.00 & 26.77  & 48.13 &
        & 48.04  &  2.47 & AGS\\
        &-70.04 &-62.67 &-1.10  &(1.195) & 23.09 & 22.13  & 40.35
        &(-43.29) &  45.25 &  2.20 & FY \\ \hline
\end{tabular}
\end{center}
\end{table}

\section{Conclusion}

In this work
we have presented the low energy $n-\nuc{3}{H}$ elastic cross sections
calculated by solving the 4-nucleon problem with realistic NN interactions
and by using three different approaches.

\bigskip
The results of Hyperspherical Harmonics and Faddeev-Yakubovsky
methods are converged to 1\% accuracy and  are in close agreement
with each other. AV18 and Nijm-II nucleon-nucleon potentials
underestimate the experimental cross section by 10\% near the
resonance peak. This disagreement is not corrected by UIX three-nucleon
forces, whose effect in this energy region is small and tends to
further decrease the theoretical value upto 15\%
~\cite{LC_FB17_03,LC_Trento_03,CarbLaz_PRC04}.

\bigskip
The results of AGS equations are closer to the experimental data.
However they have been obtained using a rank-one approximation to
the two-body t-matrix. This approximation, which was found to
provide excellent results for nuclear systems dominated by S-waves
(3N problem or positive parity 4N states), seems to provide an
excess of attraction in the negative parity states of $n-\nuc{3}{H}$
system, where P-waves turn out to be important. We believe that
accurate treatment of these resonant states requires the inclusion
of higher rank terms.

\bigskip

The results presented here concern local NN potentials.
However similar conclusions were found in previous
works~\cite{CarbLaz_PRC04} using non local ones
(Bonn~\cite{MSS96} and Doleschall~\cite{Dea03}).
This indicates a serious difficulty of the existing NN force models in
describing the simplest nuclear resonance, i.e. the $n-\nuc{3}{H}$
system. This difficulty can hardly be solved by the inclusion of a
standard type of 3NF, used to reproduce the few-nucleon
binding energies. Their origin could rather lie either in the NN
forces themselves, or in the presence of 3NF of unknown type.

\bigskip
{\bf Acknowledgment:}
The numerical calculations have been performed
with the CCRT (CEA) and IDRIS (CNRS).
We thanks the staff members of both computer centers for their constant
help. The work of (ACF) was supported in part by grant
POCTI/FNU/37280/2001.



\begin{thebibliography}{99}


\bibitem{CCG_98} F.~Ciesielski, J.~Carbonell, C.~Gignoux,
                 Nucl. Phys. A {\bf {631}} 653c (1998).
\bibitem{CC_98} F. Ciesielski and J. Carbonell,
               Phys. Rev. C {\bf 58},  58 (1998).
\bibitem{CCG_99} F. Ciesielski, J. Carbonell and C. Gignoux,
 Phys. Lett. B {\bf{447}}, 199 (1999).
\bibitem{CCGF_98}  F. Ciesielski, J. Carbonell, C. Gignoux, A. Fonseca,
                   Contribution to the XVI European Conference on
                              Few-Body Physics, Autrans 1998.
\bibitem{Viv98}   M. Viviani, S. Rosati and A. Kievsky,
                  Phys. Rev. Lett. {\bf 81}, 1580 (1998).
\bibitem{AF_99}  A.C.  Fonseca, Phys. Rev. Lett. {\bf 83}, 4021
(1999).
\bibitem{NKG00} A. Nogga, H. Kamada, W. Gl\"ockle,
  Phys. Rev. Lett. {\bf 85}, 944 (2000).
\bibitem{Viv01}  M. Viviani, A. Kievsky, S. Rosati,
                 E. A. George and L. D. Knutson,
                 Phys. Rev. Lett. {\bf 86}, 3739 (2001).
\bibitem{AF_02}  A.C.  Fonseca, G. Hale and J. Haidenbauer, Few Body
Systems {\bf 31}, 139 (2002).
\bibitem{TWH92} D. R. Tilley, H. R. Weller, and G. M. Hale,
                Nucl. Phys. A {\bf 541}, 1 (1992).
\bibitem{Kea01} H. Kamada {\it et al}.,
               Phys. Rev. C {\bf 64}, 044001 (2001).
\bibitem{AV18+} B. S. Pudliner, V.R. Pandharipande,J. Carlson, S.C.Pieper and R.B. Wiringa,
               Phys. Rev. C {\bf 56}, 1720 (1997).
\bibitem{KG92} H. Kamada, W. Gl\"ockle, Nucl. Phys. A {\bf 548}, 205
(1992); W. Gl\"ockle, H. Kamada, Phys. Rev. Lett. {\bf 71}, 971  (1993).
\bibitem{KKF89} H. Kameyama, M. Kamimura, and Y. Fukushima, Phys. Rev. C {\bf
40}, 974 (1989).
\bibitem{KK90} M. Kamimura and H. Kameyama,
 Nucl. Phys. A {\bf 508}, 17c (1990).
\bibitem{VS95}  K. Varga and Y. Suzuki, Phys. Rev. C {\bf 52}, 2885
  (1995).
\bibitem{SV98}  Y. Suzuki and K. Varga, {\sl Stochastic variational approach
to quantum mechanical few-body problems}, Springer-Verlag, 1998.
\bibitem{VKR_04}  M. Viviani, A. Kievsky, and S. Rosati, nucl-th/0408019
\bibitem{VKR_95}   M. Viviani, A. Kievsky, and S. Rosati,
   Few-Body Systems {\bf 18}, 25 (1995).
\bibitem{C88} J. Carlson, Phys. Rev. C {\bf 38}, 1879 (1988).
\bibitem{Wea00} R. B. Wiringa, S. C. Pieper, J. Carlson, and
 V. R. Pandharipande,  Phys. Rev. C {\bf 62}, 014001 (2000).
\bibitem{NB99} P. Navr\'atil and B. R. Barrett, Phys. Rev. C {\bf 59},
  1906 (1999).
\bibitem{NVB00} P. Navr\'atil, J. P. Vary and B. R. Barrett,
  Phys. Rev. Lett. {\bf 84}, 5728 (2000); Phys. Rev. C {\bf 62}, 054311 (2000).
\bibitem{BLO00} N. Barnea, W. Leidemann, G. Orlandini,
  Phys. Rev. C {\bf  61}, 054001 (2000).
\bibitem{Nogga_alpha} A. Nogga, H. Kamada, W. Gl\"{o}ckle and B.R. Barrett,
Phys. Rev. \textbf{C 65} (2002) 054003
\bibitem{Y67} O. Yakubovsky, Sov. J. Nucl. Phys. {\bf 5}, 937 (1967).
\bibitem{PHH01} B. Pfitzinger, H. M. Hofmann and G. M. Hale,
                Phys. Rev. C {\bf 64}, 044003 (2001).
\bibitem{GSAGS_67} P. Grassberger and W. Sandhas, Nucl. Phys.
B {\bf{2}}, (1967) 181; E.O. Alt, P. Grassberger, W. Sandhas,
Phys. Rev. C {\bf{1}}, 85 (1970); ibdem JINR Report No.
E4-6688 (1972).
\bibitem{Fred_97}  F. Ciesielski, {\em Th\`ese Univ. J. Fourier (Grenoble)\/}
(1997).
\bibitem{LC_FB17_03} R. Lazauskas, J. Carbonell,
                  Nucl. Phys. A {\bf 737} S79 (2004).
\bibitem{LC_Trento_03} R. Lazauskas, J. Carbonell,
      Few-Body Systems {\bf 34}, 105 (2004).
\bibitem{AV18} R. B. Wiringa, V. G. J. Stoks, and R. Schiavilla,
               Phys. Rev. C {\bf 51}, 38 (1995).
\bibitem{NIJ_93}   V.G. J. Stoks, R. A. M. Klomp, C. P. F. Terheggen, J. J. de
Swart, Phys. Rev. C {\bf 49}, 2950 (1994).
\bibitem{AF_89.1}  A.C.  Fonseca, Phys. Rev. C {\bf 40}, 1390 (1989).
\bibitem{HS_81} H. Haberzettl and  W. Sandhas, Phys.
Rev. C {\bf{24}}, 359 (1981).
\bibitem{FS_76} A.C. Fonseca and  P.E. Shanley, Phys.
Rev. C {\bf{14}}, 1343 (1976).
\bibitem{EST_73} D.J. Ernst, C.M. Shakin and R.M. Thaler,
Phys. Rev. C {\bf{8}}, 46 (1973).
\bibitem{SMcGF_79} S. Sofianos, N.J. McGurk and  H. Fiedeldey,
Nucl. Phys. A {\bf{318}}, 295 (1979).
\bibitem{KRV94} A. Kievsky, S. Rosati, and M. Viviani,
                Nucl. Phys. {\bf A577},  511 (1994).

\bibitem{F83} M. Fabre de la Ripelle, Ann. Phys. (N.Y.) {\bf 147}, 281
             (1983).
\bibitem{KMRV96} A. Kievsky, L.E. Marcucci, S. Rosati and M. Viviani,
                 Few-Body Systems  {\bf 22}, 1 (1997).
\bibitem{VKRip}   M. Viviani, A. Kievsky, and S. Rosati,
   in preparation.

\bibitem{RL_Thesis} R. Lazauskas Thesis, UJF Grenoble (2003),
http://tel.ccsd.cnrs.fr/documents/archives0/00/00/41/78/
\bibitem{CarbLaz_PRC04} R. Lazauskas, J. Carbonell, Phys. Rev. C {\bf 70}, 044002 (2004).
\bibitem{PBS_80}   T.W. Phillips, B. L. Berman, J. D. Seagrave,
 Phys. Rev. C {\bf 22} 384 (1980).
\bibitem{KH86} Y. Koike and J. Haidenbauer, Nucl. Phys. {\bf A463},
               365c (1987).

\bibitem{WGC88} H. Witala, W. Gl\"ockle, and T. Cornelius,
                Nucl. Phys. {\bf A491}, 157 (1988).

\bibitem{KRTV96} A. Kievsky  {\it et al.},
                 Nucl. Phys. {\bf A607}, 402 (1996).

\bibitem{WG91} H. Witala and  W. Gl\"ockle,
               Nucl. Phys. {\bf A528}, 48 (1991).

\bibitem{K99} A. Kievsky, Phys. Rev. C {\bf 60}, 034001 (1999).

\bibitem{CS01} L. Canton and W. Schadow,
                Phys. Rev. C {\bf 64}, 031001(R) (2001).

\bibitem{LACG_59} Los Alamos Physics and Cryogenics Groups, Nucl. Phys. 12, 291 (1959).

\bibitem{MSS96} R. Machleidt, F. Sammarruca, and Y. Song,
                Phys. Rev. C {\bf 53}, R1483  (1996).

\bibitem{Dea03} P. Doleschall, I. Borbely, Z.Papp, W. Plessas,
                Phys. Rev. C {\bf 67}, 064005 (2003).


\end{thebibliography}
\end{document}